\documentclass[sigconf]{acmart} 

\AtBeginDocument{%
  }

\renewcommand\footnotetextcopyrightpermission[1]{}
\usepackage{color}
\usepackage{graphicx}
\usepackage{subfigure} 
\usepackage{booktabs} 
\usepackage{multirow}
\usepackage{placeins}
\usepackage{float}
\usepackage{adjustbox}
\usepackage{tabularx}
\usepackage{makecell}

\newcommand{\nccnews}{AnonymizedNews}

\begin{document}

\title{The Role of Human Creativity in the Presence of AI Creativity Tools at Work: A Case Study on AI-Driven Content Transformation in Journalism}
\renewcommand{\shorttitle}{The Role of Human Creativity}

\author{Sitong Wang}
\affiliation{
  \institution{Columbia University}
  \city{New York}
  \state{NY}
  \country{USA}
}
\email{sw3504@columbia.edu}

\author{Jocelyn McKinnon-Crowley}
\affiliation{
  \institution{Syracuse University}
  \city{Syracuse}
  \state{NY}
  \country{USA}
}
\email{jmckin02@syr.edu}

\author{Tao Long}
\affiliation{
  \institution{Columbia University}
  \city{New York}
  \state{NY}
  \country{USA}
}
\email{long@cs.columbia.edu}

\author{Kian Loong Lua}
\affiliation{
  \institution{Syracuse University}
  \city{Syracuse}
  \state{NY}
  \country{USA}
}
\email{klua01@syr.edu}

\author{Keren Henderson}
\affiliation{
  \institution{Syracuse University}
  \city{Syracuse}
  \state{NY}
  \country{USA}
}
\email{khenders@syr.edu}

\author{Kevin Crowston}
\affiliation{
  \institution{Syracuse University}
  \city{Syracuse}
  \state{NY}
  \country{USA}
}
\email{crowston@g.syr.edu}

\author{Jeffrey V. Nickerson}
\affiliation{
  \institution{Stevens Institute of Technology}
  \city{Hoboken}
  \state{NJ}
  \country{USA}
}
\email{jnickers@stevens.edu}

\author{Mark Hansen}
\affiliation{
  \institution{Columbia University}
  \city{New York}
  \state{NY}
  \country{USA}
}
\email{mh3287@columbia.edu}

\author{Lydia B. Chilton}
\affiliation{
  \institution{Columbia University}
  \city{New York}
  \state{NY}
  \country{USA}
}
\email{chilton@cs.columbia.edu}

\begin{abstract}
As AI becomes more capable, it is unclear how human creativity will remain essential in jobs that incorporate AI. 
We conducted a 14-week study of a student newsroom using an AI tool to convert web articles into social media videos. 
Most creators treated the tool as a creative springboard, not as a completion mechanism.  
They edited the AI outputs. 
The tool enabled the team to publish successful content that received over 500,000 views. 
Human creativity remained essential: after AI produced templated outputs, creators took ownership of the task, injecting their own creativity, especially when AI failed to create appropriate content. 
AI was initially seen as an authority, due to creators' lack of experience, but they ultimately learned to assert their own authority.
\end{abstract}



\maketitle
\pagestyle{plain} 

\section{Introduction}
The integration of AI into creative processes is revolutionizing content creation, particularly in the fast-paced environment of newsrooms. 
As generative AI tools become increasingly prevalent, understanding the intricate interactions between these technologies and human creativity is crucial. 
What is the role of human creativity in the presence of AI creativity tools in the workplace? 
Does the role of human creativity change over time as people gain familiarity with AI tools?
Although AI has some capacity for creation, creativity is a complex cognitive process involving many skills, including divergent and convergent thinking, framing, analysis, critique, audience understanding, and iteration. 
AI may be able to assist with some or parts of these processes, but human creativity is likely to still play a role in the creative process.

To explore the role of human creativity in the presence of AI, this paper presents a 14-week study conducted within the social media team of a university journalism department's newsroom, focusing on content creation for the TikTok platform. 
The team included five social media creators (all students), one editor (a faculty member), and two market researchers (research assistants), embodying a typical real-world newsroom structure.
They used an AI tool designed to help creators transform web articles into engaging short video scripts and storyboards. 
Using the tool, the journalists read, selected, regenerated, and edited several possible AI-generated scripts before shooting and editing the videos themselves. 
Each journalist aimed to produce one new video a week and iterate on previous videos. 
The newsroom met weekly for the editor to review videos, request changes, and decide if a video was ready to publish. 
After 10 weeks, the market researchers conducted 42 interviews with members of the target audience to gather suggestions for improvement. 

\begin{figure*}[t]
\centering
\includegraphics[width=0.94\textwidth]{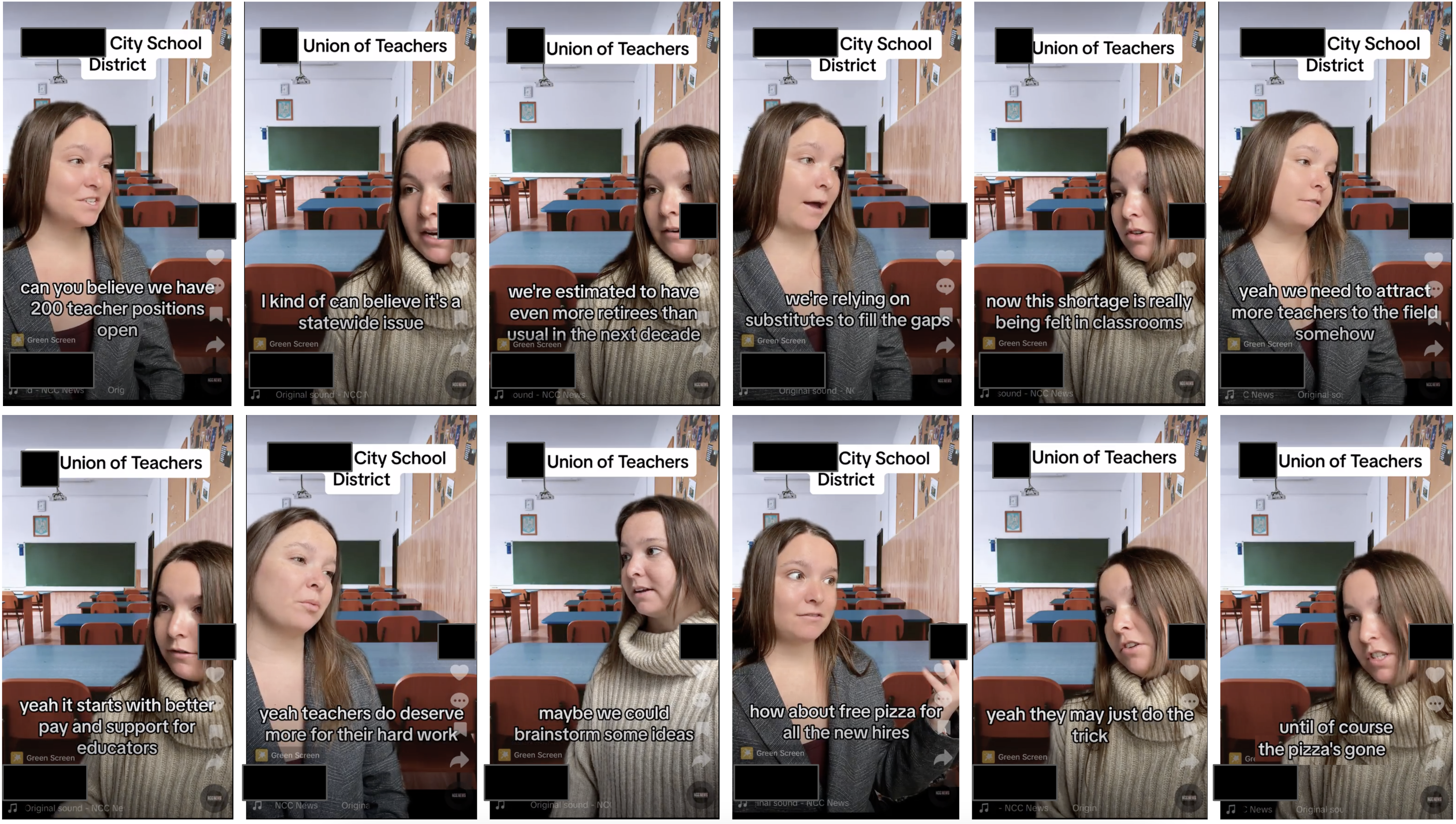}
\caption{
A news reel made during the student study covering local news about a teacher shortage. 
}
\label{fig:rf2-ex-schools}
\end{figure*}

Overall, the team was successful in publishing 16 videos in 14 weeks and amassing over 500,000 views. 
AI was far from perfect---creators had to edit ``about half'' of the AI scripts for accuracy and clarity. 
However, \textbf{AI was helpful both as scaffolding when learning the task, and as a creative springboard}~\cite{shneiderman2007creativity} \textbf{even after they understood the task.}  
Additionally, the tool was not strictly necessary---creators opted not to use the tool when it was not working well for a story or if they had a specific creative inspiration.
In addition to correcting AI, creators also intervened to inject their own creativity in many parts of the process.
\textbf{Whereas AI could only produce outputs, the creators took ownership over the task, using their judgment and creativity to maximize its value to audiences}, rather than just adapting a story to a template as AI did. 
The creators' relationship with AI changed over the course of the 14-week study.
Initially, they relied heavily on the AI's outputs---even accepting its blatant errors without question. 
However, through critical reflection and iteration, creators were able to reassert themselves.
\textbf{AI was initially seen as ``an authority'' on news reels, due to creators' lack of experience and AI's mystique of intelligence. 
But ultimately, users were better able to balance AI's authority and their own.}

\section{Background and Related Work}

\begin{figure*}[t]
\centering
\includegraphics[width=0.96\textwidth]{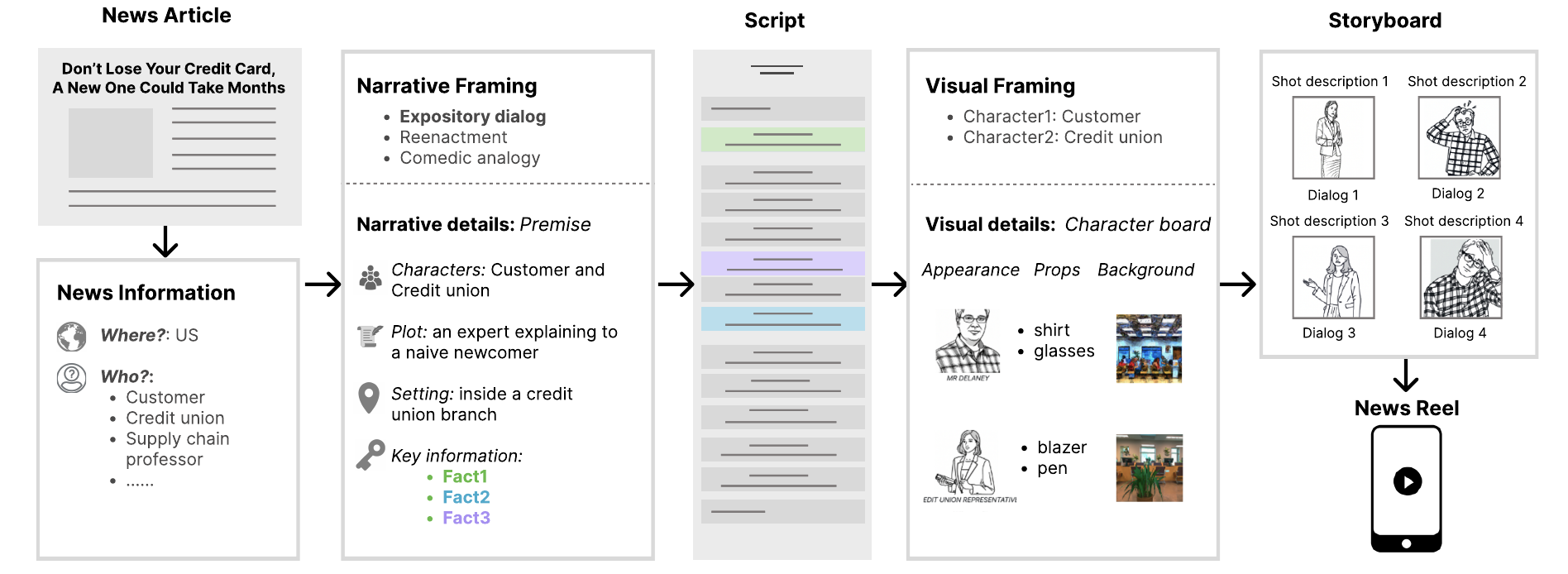}
\caption{
ReelFramer is a human-AI co-creative system that supports journalists in creating news reels from print articles.
Users input the text of a news article, and accept or reject AI suggestions for news information, narrative framing elements, script, visual framing, and a storyboard. 
}
\label{fig:teaser}
\end{figure*}

\subsection{Studies of AI in a Work Context}
Recent studies have mapped out the significant opportunities that generative AI brings to various industries, highlighting its potential to enhance productivity for a wide range of workspace tasks~\cite{brynjolfsson2023generative, scienceproductivity}. 
It has shown promise in replacing both cognitive routine work (often considered drudgery) and cognitive non-routine work, including creative tasks~\cite{bouschery2023augmenting}, professional writing tasks~\cite{scienceproductivity}, professional software development~\cite{Cui_2024, peng2023impact}, and professional analysis work~\cite{dell2023navigating}.

However, various studies also express reservations about AI's ability to fully replace human workers. 
A study of creativity support tools demonstrates that while the tools are helpful for specific tasks, they cannot solve problems entirely on their own~\cite{tao2023ai}---especially for tasks involving core human skills, even in the long run~\cite{long2024novelty}. 
AI support can even be problematic. 
\cite{dell2023navigating} shows that for difficult tasks, subjects using ChatGPT were about 20\% less likely to make the correct recommendation, as many were led astray by its convincing but incorrect analysis. 
Even when AI is correct, users face challenges incorporating AI results \cite{Kim2024}. 
These limitations suggest that AI might not completely reshape entire industries but rather impact certain aspects of how people work and what they focus on~\cite{randazzo2024cyborgs}. 

\subsection{AI for Creativity Support}
The use of AI for creativity support, particularly to assist writing, has expanded rapidly in recent years, with applications across various domains, including journalism~\cite{opal,reelframer,anglekindling}. 
AI writing assistants can provide scaffolding for writers by supporting different cognitive stages of the writing process~\cite{flower1981cognitive}, including framing~\cite{hui2023lettersmith, reelframer}, idea generation \cite{clark2018creative, anglekindling} to drafting~\cite{sparks, VISAR, tweetorial_hook, menon2024moodsmith} and revision \cite{sparks, wu2019design}. 
\cite{lee2024design} provides a comprehensive design space for intelligent and interactive writing assistants that emphasizes the importance of considering task, user, technology, interaction, and ecosystem aspects when using AI to support writers. 
However, using AI in creativity also has risks; overreliance~\cite{overreliance_ibm, overreliance_facct, overreliance_examples, overreliance_difficulty} on AI can cause people to accept poor AI answers, and AI ideation can actually reduce diversity across people~\cite{kreminski_homogenity}.
Broadly, AI can be seen not as a solution, but as a design material that needs to be understood by human creators \cite{ai_design_material_CHI22, eytan_design_material}.

A major creative application of AI lies in content retargeting and transformation—editing and repurposing text, image, video, or audio for new audiences, platforms, or goals.
The key is to find the fundamental message to preserve while adapting the content into the new format.
Previous works have explored how to support people with content transformation tasks in long-form and short-form video creation~\cite{truong2021automatic,chi2021automatic,chi2022synthesis,podreels,Rope_wang}.
In particular, short-form content requires restructuring narratives to fit new constraints and audience expectations~\cite{kim2024unlocking}. 

\subsection{News Reels}
News reels help outlets reach younger viewers who favor TikTok and Instagram~\cite{newman_news_tt_reuters_oxford, newman_digital_news_2022}. 
Whereas traditional journalistic narratives are serious and fact-driven, news reels often use a playful tone that balances between information and entertainment,  which meets expectations for social media videos~\cite{newman_news_tt_reuters_oxford}. 

A common way to reframe news for reels is to create a role-play. An actor/journalist creates a dialogue between two characters related to the news story and acts out the perspective of each character. 
For example, in a Washington Post reel, a character representing Hurricane Ian explains to a roommate character why he is “visiting” Florida and all the trouble he might cause there. 
Figure \ref{fig:rf2-ex-schools} contains an example of a news reel made in our newsroom study with a dialogue between a School District character and a Teachers Union character acting out the news of a teacher shortage. 
As is typical of news reels, the creator is playing both roles.

\subsection{ReelFramer AI Tool}
ReelFramer~\cite{reelframer} is a creativity support system that uses generative text and image AI to scaffold the process of creating a script and storyboard for a news reel. 
The user starts by inputting the news article (see Figure \ref{fig:teaser}). 
The system uses a large language model (LLM), GPT-4, to extract locations, people, and activities from the article. 
The user then selects one of the three narrative framings, which contain a different balance of information and entertainment (expository dialogue, reenactment, or comedic analogy). 
The system then uses the LLM to suggest foundational details like characters, plot, setting, and key information, which is called the premise in scriptwriting~\cite{batty2017script}.
The system is iterative and built for iteration; users can accept, regenerate, or edit any of the AI generations.
Once the user is happy with the script, the system provides visual framing to explore the characters’ visual design.
It generates a character board that contains visual details of the costumes, props, and backgrounds to distinguish the characters using a text-to-image model, DALL-E 2. 
Based on the character board, it generates a storyboard that shows suggested emotions, actions, and dialogues for each character. 
Users can then film the reels based on these materials.

\section{Student Newsroom Study}
To understand the role of human creativity in the presence of an AI support tool, we present a 14-week study of how a student newsroom produces news reels using the ReelFramer AI tool. 

\subsection{Methods}
In the student newsroom study, researchers recruited advanced students from a university journalism program that runs a TV and print journalism media outlet called \nccnews. 
As part of their existing program, the students report local area news in print and on television with faculty guidance and approval. 
Before the study, \nccnews~had a small TikTok channel, with fewer than 50 posts and no posts getting more than 3000 views. 
Simulated newsroom environments are common in journalism education spaces~\cite{steel2007experiential} as part of practical and experiential learning with both procedural (environment) and factual (task) authenticity~\cite{chen2001pedagogy}.  
Choosing students in a university newsroom was fairly representative of professional newsrooms because they often hire recent graduates to run social media, due to their familiarity with social media, and the demographic they aspire to reach~\cite{newman_digital_news_2022}.

Recruitment emails were sent to senior journalism classes offering an opportunity to learn about creating news reels with AI in a newsroom environment and to regularly publish videos under editorial guidance. 
Like an internship, this would be a paid opportunity.
Six student journalists attended the first training session, and five participated in the full 14-week newsroom simulation.  
The size of this sample, along with attrition, is expected for longitudinal qualitative research~\cite{saldana2003longitudinal}.

In this simulated newsroom environment, we had one researcher acting as editor, two researchers acting as market researchers, and five students given news stories each week to convert to reels, as if they were part of a digital news desk. 
The news stories were ones the newsroom had previously reported. 
In weekly meetings, researchers observed students ~\cite{emerson2011writing}, and the editor interacted with the student journalists to offer the support an editor would provide~\cite{seim2024participant} in a standard newsroom. 
Student journalists were regularly asked about their use or nonuse of the tool.
The editor led group feedback discussions for every video and gave specific feedback for iteration. After approval by the editor, newsreels were published to the \nccnews~TikTok channel.

\subsection{Findings: Tool Use and Disuse}

Overall, the student group made 26 news reels, 16 of which were published by the editor. 
The 10 unpublished were still being iterated at the end of the session or had been abandoned.
Altogether, the 16 videos published by the team garnered over 500,000 views. 
Four of the five student journalists chose to use the tool for all or most of their videos. 
13 of 16 published videos used ReelFramer in a significant manner to help achieve an output the editor deemed publishable. 

Creators choose to use the AI tool a majority of the time---for 20 of 26 reels created. To understand what, if any, benefits the tool had, we analyzed why users used the tool and why they chose not to use the tool.

\subsubsection{Reasons for Usage: Framing and Drafting Scripts}

The users repeatedly mentioned the benefit of two of the steps, each supported a different cognitive process.
Users mentioned not using the storyboarding features, as it was not cognitively demanding.

1. \textbf{Narrative framing.} 
They all used the narrative framing features of ReelFramer to find the characters, premise, and key news facts in the web article that they could use to transform into a role-play script.  
They described this as a cognitively demanding task to ``see the story'' from a different narrative frame. ReelFramers' suggestions were typically accurate, creating a reliable premise for a script.

2. \textbf{Drafting scripts from premises.} 
They used the scriptwriting features of ReelFramer to initiate a draft of the script. Although generating the premise was helpful, there was another cognitive challenge of expanding that to a script. Translating raw ideas into fluent text is a complex writing task that requires perspective-taking, thematic grouping and sequencing of information, and managing cohesion and flow ~cite{flower1981cognitive}.

However, the users consistently reported editing ``about half'' of the AI-generated script. 
They found they had to improve many of the writing mechanisms, such as taking out awkward phrases, increasing clarity, or removing lines with too little content. 
Overall, it was a helpful starting point.

Notably, those who used the tool only used it for the narrative framing and scripting steps. None of the students used the storyboarding features, as they did not find this part of the task cognitively demanding; they said they could ``see'' how the story would be filmed.  

\subsubsection{Reasons (and Risks) of Disuse}

Creators chose not to use the AI tool for 6 of the 26 reels. In some cases, this was reasonable; in others, it led to problems.

1. \textbf{Strong understanding of the story (2).} Two creators skipped ReelFramer when adapting stories they had originally reported. Having done the initial reporting made it easier to reframe the story into a role-play narrative without the tool.

2. \textbf{Task deemed easy (2).} P5 used ReelFramer once, then stopped, believing he understood the framework. While his reels had creative additions and were entertaining, they were not publishable as they all lacked key news information. This suggests ReelFramer could help as a scaffolding tool, even for those who do not want or need it.

3. \textbf{Independent inspiration (1).} P2 had an independent inspiration for his own framing (a monologue rather than a role-play), which he executed by himself. The editor liked the inspiration, but found the monologue format confusing. He iterated to bring back the role-play style framing and was ultimately successful in publishing the video. Independent inspiration could become more scaffolding.

4. \textbf{Tone-deaf AI (1).} P1 skipped the tool when its ``fun'' and ``quirky'' tone did not match the seriousness of the topic---domestic violence. 
She took inspiration from the format she had learned from the AI tool and wrote her own script, and the video was successfully published. For her future reels, she returned to the tool due to its cognitive support.

\subsubsection{Editors' Reflections Over Time}

In general, the tool made two cognitively challenging tasks easier---it was not always necessary, but it was helpful. The editor observed that in the beginning, the tool served as a teaching instrument to \textbf{scaffold the process} during the learning phase. This was particularly helpful for reframing---whereas journalists are taught to write stories in a style that is typically serious, and written from a more objective point of view, the reframing steps that help journalists ``see'' a story in this new way, also provided the scaffolding to help them find these transformations for themselves.

Additionally,  when a user chose not to use the tool, their reels had fundamental problems such as not balancing information with entertainment---the core challenge the tool supports. For some users, positioning the tool only as a scaffolding tool for learning could help them understand the complexities of the task before setting out on their own. 

There were incidents when the tool failed, and it was clear that the creators could make new reels without it. However, 4 of the 5 users still chose to use the tool. They treated it as a \textbf{``creative springboard''}---rather than a replacement for their creativity, particularly when it came to drafting scripts based on the premise. Finding the words to use in a script is hard, and it is easier to edit a script than generate one from scratch. 

\subsection{Findings: Role of Human Creativity}
\label{sec:creativity}

\begin{figure}
\centering
\includegraphics[width=1\columnwidth]{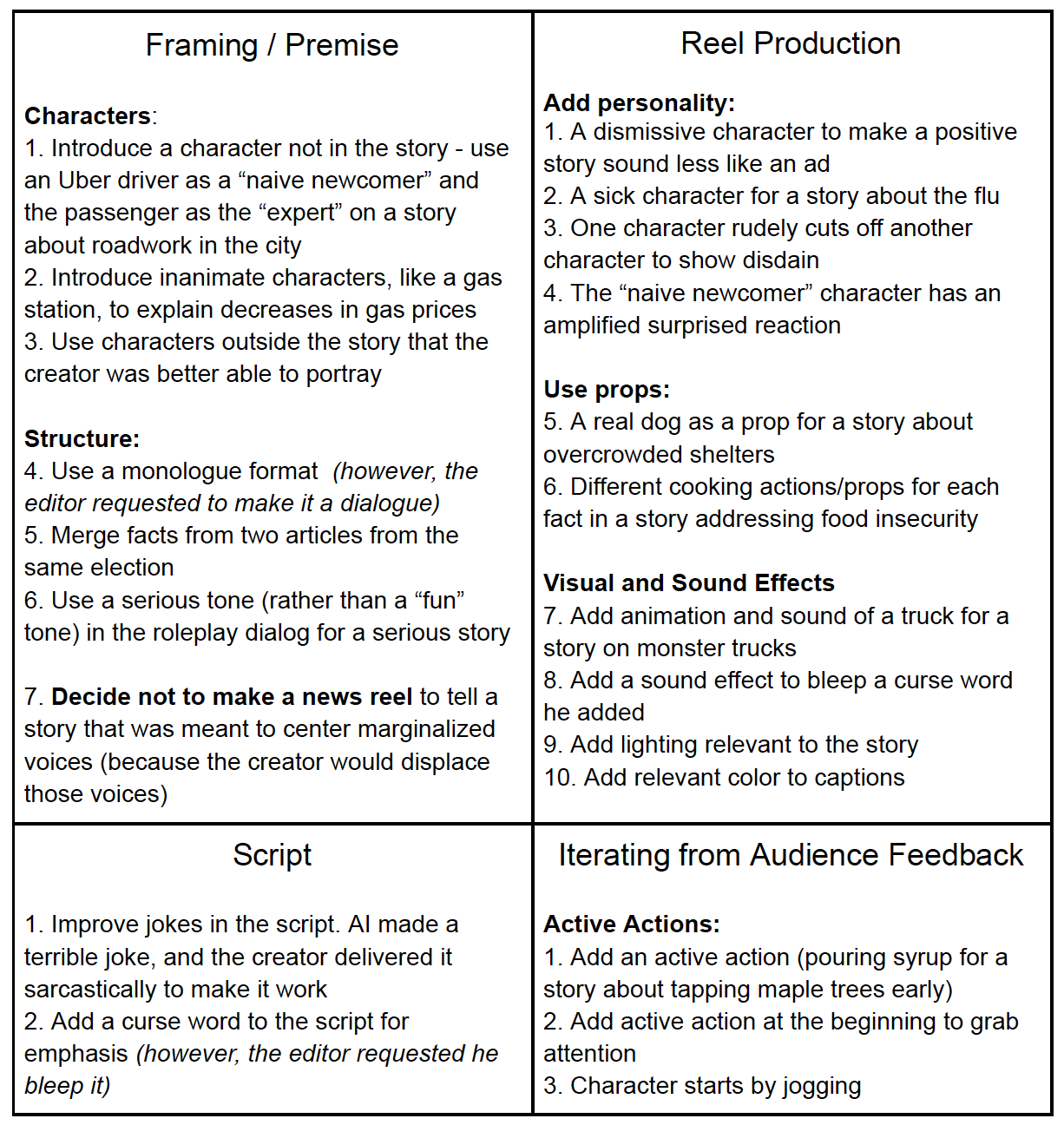}
\caption{Creative additions made by users.}
\label{fig:creative_adaptions}
\end{figure}

To understand when and how creators chose to use their own creativity rather than relying on AI, we analyzed all their creative additions made to news reels created during the study that ReelFramer did not (or could not) make.
This analysis is based on a list of creative additions noted by the editor every week during feedback sessions. This includes both published and unpublished reels. 
Some changes, like correcting an awkward AI phase, or adding captions, or correcting facts, were not deemed creative, as they were fairly typical copy editing changes. However, changes like introducing new types of characters, or adding new elements (actions, personality, animations, etc) were deemed creative because they significantly deviated from the outputs that AI could have produced.

Across the 26 videos made by the students, the editors noted 22 instances where the creator made substantial creative additions (see Figure \ref{fig:creative_adaptions}). The editor also denoted the four parts of the creative process where they were made. 

1. \textbf{Framing or Premise (7)}
Overall, 7 creative additions were made in the framing or premise of the news reel. Whereas framing and premise are supported by the AI tool, users choose to make improvements to this area based on either inspiration or need. This includes finding different characters to base the role play around, introducing new structures for the narrative, and deciding not to make a news reel at all.

Whereas the tool limited itself to using characters directly taken from the news story, the users knew that this was not strictly necessary for TikTok videos. They could use inanimate objects or tangential characters if they were more fun or easier to depict in the video. The tool was also artificially limited in its structures, but users needed to change that---this included merging facts from two news stories to improve the narrative (rather than working with only one story as the tool assumes). The tool also assumes the tone should be quirky and fun, but for a story about a serious subject of domestic violence, the creator realized the ``fun'' tone was inappropriate, and decided to make a role-play but with a serious tone. This was a large departure from the tool---and the typical newsreel format---but it was necessary, and the reel got the most views by far.

Users also needed to exercise creative and professional judgment to determine when a news reel format would be inappropriate.
The article was titled ``Women’s Club Basketball Discusses What Black History Month Means to Them''. 
None of the journalists identified as both Black and female, and thus role-play narrative felt socially inappropriate. 
Additionally, they did not want to reframe the narrative around a different set of characters because it would displace the voices of the people it was trying to center. 
Ultimately, they decided that not making the reel was the right decision.

2. \textbf{Script (2)}
Overall, most users appreciated AI most for its ability to produce scripts. All AI-generated scripts needed editing, but sometimes those edits had to be creative. AI would often add terrible jokes to scripts and users had to find a way to improve the humor---sometimes they deliver the AI joke sarcastically, to make it funny (which is not how it was intended). Sometimes, lines were dull and needed more energy. One user added a curse word, which the editor insisted he bleep, but it still brought more energy and was funny. In both cases, users brought their own sense of humor to the script.

3. \textbf{Reel Production (10)}
A major area where users added creativity was in the production of the reel---acting, props, lighting, sound effects, and post-production like actions and animation. This is an area that the tool did not support well, and that users found there were many ways to enhance the quality of reels---as well as bring in their own knowledge and creativity. Adding a strong personality to the characters was a creative decision that made a big difference. AI scripts had a negative tendency to sound too positive or even promotional. But creators could take an enthusiastic character and make them sarcastic, or have them be rudely cut off by another character to undermine the overzealous positivity. Giving characters amplified emotions (such as being very surprised, or very miserable when sick) also gave videos more life.

Using props well also made a big difference, and creators used what they had on hand---including a real dog for a story about animal shelters, or cooking utensils for a story about food insecurity. AI could not have known what users would have, so its suggestions for props were weak, but users reacted to their environment to add drama through prop usage. Additionally, CapCut software makes post-production effects like animations, sound effects, lighting filters, and caption colors very easy. Users often found creative ways to exchange the message through this type of post-production.

4. \textbf{Iterating from Audience Feedback (3)}
In the last week, creators added three more actions to videos---pouring maple syrup, jogging, and unwrapping food. Each of these was highly relevant to the story and made the videos more fun and eye-catching.

Overall, users made significant creative additions to their reels. This included adding to processes the tool supported, such as framing and scripting, as well as places where AI left off---such as acting, visual effects, and props, or iterating based on user feedback. When making these additions, users often injected their own knowledge and taste, like their own sense of humor, their acting ability, their knowledge of post-production techniques, and the props they had on hand.
Moreover, this level of creativity indicates that users were not passively accepting AI's creativity. They actively inserted themselves in the process based on need (when AI failed) or inspiration (when they had independent ideas---often that were better than AI's suggestions).

\subsubsection{Editors' Reflections Over Time}

To the editor, it was clear that the creators were heavily invested in producing high-quality videos. Whereas AI could produce many outputs, the creators took ownership over the entire process. They critiqued outputs, considered the audience, adjusted based on what was appropriate and available, and generally tried to maximize the creative potential of each story, within the time constraints of their deadline. In stark contrast, AI was mostly following a template and executing a task as directed. The creators' sense of ownership and holistic responsibility for the quality of the product drove their continued creativity. 

\subsection{Overcoming AI Errors}

\begin{table}[ht]
\centering
\begin{tabularx}{\columnwidth}{|>{\raggedright\arraybackslash}X|>{\raggedright\arraybackslash}X|}
\hline
\multicolumn{2}{|c|}{\textbf{Major Issues (15)}} \\ \hline
\textbf{AI Attributed (11)} & \textbf{Human Attributed (4)} \\ \hline
too positive/promotional (4)\newline 
too long (2)\newline 
too few facts (2)\newline 
unclear narrative (1) \newline
too wordy (1) \newline
abrupt intro (1)
&
slow paced (2)\newline confusing format -- monologue (1)\newline biased narrative (1) \\
\hline
\end{tabularx}
\caption{Major Issues: AI vs Human Attributed in the student newsroom study.}
\label{tab:major_issues}
\end{table}

\begin{table}[ht]
\centering
\begin{tabularx}{\columnwidth}{|>{\raggedright\arraybackslash}X|>{\raggedright\arraybackslash}X|}
\hline
\multicolumn{2}{|c|}{\textbf{Minor Issues (19)}} \\ \hline
\textbf{AI Attributed (15)} & \textbf{Human Attributed (4)} \\ \hline
awkward AI phrase (4)\newline
confusing character names (3)\newline
ending lacks polish (2)\newline
too positive/promotional (1)\newline
strays from news (1)\newline
tone shift (1)\newline
unclear/missing fact (2)\newline
unnecessary dialogue (1)
&
labels too long (1)\newline
added curse word (1)\newline
add relevancy to the present (1)\newline
missing facts (1)
\\ \hline
\end{tabularx}
\caption{Minor Issues: AI vs Human Attributed in the student newsroom study.}
\label{minor_issues}
\end{table}

AI is widely known to produce errors---as do humans. 
The newsroom in this study had an editor who was in charge of quality control. 
This is typical of all newsrooms. 
Similar to typical editorial newsroom dynamics, creators submit drafts, receive critiques and edits from the editor, and have to iterate to produce a new draft. 
Iteration continues until the editor decides to publish the piece or decides not to run it. 

To understand the nature of the edits, we tracked two things: the magnitude of the error (major or minor) and the attribution of the error (whether the error was introduced by the AI or the human journalist). 

For error magnitude, major edits are those that required reshooting the entire news reel; minor edits required reshooting only part of the video (e.g., a localized scene or dialog edits).
We also tracked whether the errors were introduced by AI or by the creators, as judged by the editor from discussions with the creator.
If the script contained incorrect information, and the script was drafted by AI and edited by a creator, the editor attributed this error to being introduced by AI. 
Although it should have been caught by a human, it might not have needed to be caught if it had not been introduced by AI. 
However, if the human writes a script with incorrect information, this error is attributed to the human.

Surprisingly, a vast majority of the errors were introduced by the AI tool. 
11 of 15 major errors, and 15 of 19 minor errors were introduced by AI (Tables \ref{tab:major_issues} and \ref{minor_issues}).
These errors were accepted by the creators, and they filmed news reels with these errors in them. 
Although it is expected that AI will make errors, and that many people using AI on a deadline might simply accept AI content at face value and not check it carefully. 
However, this case is different---creators actually had to speak the lines AI wrote, so they were fully cognizant of the dialogue they were including. 

The most common major issue was clearly attributed to the AI tool---sounding too positive. 
The AI is trying to balance entertainment and information in a ``quirky'' way, and the stories should not sound sad and serious like traditional evening news anchors.
However, AI leaned too heavily toward a positive tone, and this bias clashed with the content in different ways. 
For stories that were serious or sad (stories about homelessness, domestic violence, or climate change), the ``quirky'' tone was inappropriate for the topic. 
For lighter stories---like a small town winning an award for being charming, AI's positive tone was also a problem. 
The positive tone made the video feel like an advertisement rather than news. 
In all cases, the team had to find creative ways to fix the narratives to be more appropriate. 

The most common minor issue clearly attributed to the AI tool (and accepted by creators) was including bizarre and awkward AI phrases.
These stick out badly and are fairly localized changes to make, so the effort to change them should have been minimal. 
Phrases like \textit{``a little bird told me''} are awkward AI phrases that were inappropriate for a news reel because all news should make some effort to attribute information to sources. 
Other awkward phrases include \textit{``a comedy of wonders''} and distracting attempts at wordplay such as \textit{``a two-tiered system? Is that kind getting rewarded with two kinds of desserts?''}
When asked why they included these phrases, the typical response was ``I guess it seemed OK.''

This acceptance indicates an implicit trust in AI, even when the AI does a task in the creator's realm of expertise. 
A crucial role of human creativity is to be critical of AI outputs.
Even journalists who are trained to be critical may require additional training to learn to be critical in this new way.

\subsubsection{Editors' Reflection Over Time}

The editor noticed a general trend that, initially, users relied heavily on
the AI’s outputs---even accepting its blatant errors without question. However, through critical reflection
and iteration, creators were able to reassert themselves.
AI was initially seen as an authority on news reels,
due to the creators’ lack of experience and AI’s mystique. But ultimately, users were better able to balance AI’s authority and their own.

Through conversations with the editor and each other, creators were able to overcome this. 
It seemed to be correlated with both more group and editorial critique, as well as students having a better mental model of the strengths and weaknesses of the tool through repeated usage. 
The editor pointed out that \textit{``at some point [the students] were able to anticipate what the tool was going to suggest (to a point) because they understood how it was translating between genres [web stories to reels].''} 

In the early videos, students often ``miss the heart of the story.''
It was important for students to understand the “news value,” so the videos focused on the information rather than entertainment. 
The editor said that the students' \textit{``learning and improvement happened when they made connections between the tool and their occupational standards.''} 
Balancing between news and entertainment was one of the goals of the tool, and whereas it did help, it was still important for the creators to have their own understanding of this, so they could make appropriate edits.

\subsection{Limitations and Future Work}
This study examined one student newsroom using a specific AI tool for 14 weeks, limiting generalizability. Future work could explore longer-term effects on populations of multiple types of journalistic and AI-based experiences. In particular, professional newsrooms may face different dynamics given their tighter deadlines, higher editorial standards, and larger audiences, and other creative fields may reveal distinct patterns of human-AI collaboration.

\section{Conclusion}
This paper studied the benefits and challenges of using AI in the organizational context of a newsroom.
AI has many strengths as a creativity support tool—especially for writing—so we studied journalists, experts in the craft of writing and storytelling.
Newsrooms and other creative industries are places where AI could significantly impact work—either by changing its nature or potentially displacing it.
Overall, we found that AI was helpful both as scaffolding during the learning phase and as a creative springboard once users understood the task.
AI supported human creators but did not replace them. While it could generate outputs, it was the creators who took ownership of the task, using judgment and creativity to tailor stories for audiences, rather than merely adapting them to templates.
Despite the flexibility of generative AI, it follows instructions linearly and struggles to adapt to the nuances of individual stories or audiences.
Because AI still makes frequent mistakes and lacks contextual understanding, editorial oversight was essential to ensure quality and avoid blindly accepting flawed outputs.
Through this reflective process, creators also reasserted their expertise. While AI was initially seen as an “authority,” users ultimately learned to balance its contributions with their own.

\bibliographystyle{ACM-Reference-Format}
\bibliography{sample-base}

\end{document}